\documentclass[aps,prl,twocolumn,superscriptaddress,longbibliography]{revtex4-2}
\usepackage{braket} % For Dirac notation
\usepackage{amsmath, amssymb}
\usepackage{bm}
\usepackage[T1]{fontenc} 
\usepackage{lipsum} % For generating dummy text
\usepackage{amsmath}
\usepackage{graphicx}
\usepackage{subcaption}
\usepackage{float}
\usepackage{soul}
\usepackage{xcolor}
\sethlcolor{yellow}  % or any color you like
\newcommand{\yb}{\ensuremath{{}^{171}\mathrm{Yb}^{+}}\text{ }}
\usepackage[colorlinks=true,citecolor=blue]{hyperref}

\begin{document}

\title{Steady-States of a Single Trapped-Ion Spin Coupled to an Engineered Non-Markovian Bath}

\author{Anthony Vogliano}
\affiliation{Department of Physics and Astronomy, University of Waterloo, Waterloo, ON, N2L 3G1, Canada}

\author{Lewis Hahn}
\affiliation{Department of Physics and Astronomy, University of Waterloo, Waterloo, ON, N2L 3G1, Canada}

\author{Fabien Lefebvre}
\affiliation{Department of Physics and Astronomy, University of Waterloo, Waterloo, ON, N2L 3G1, Canada}

\author{Jingwen Zhu}
\affiliation{Department of Physics and Astronomy, University of Waterloo, Waterloo, ON, N2L 3G1, Canada}

\author{Sakshee Patil}
\affiliation{Department of Physics and Astronomy, University of Waterloo, Waterloo, ON, N2L 3G1, Canada}

\author{Mahmood Sabooni}
\affiliation{Department of Physics and Astronomy, University of Waterloo, Waterloo, ON, N2L 3G1, Canada}

\author{Zhexuan Gong}
\affiliation{Department of Physics, Colorado School of Mines, Golden, Colorado 80401, USA}

\author{Rajibul Islam}
\affiliation{Department of Physics and Astronomy, University of Waterloo, Waterloo, ON, N2L 3G1, Canada}

\date{\today}

\begin{abstract}
    Quantum simulation of open quantum systems offers a pathway towards better understanding various non-equilibrium physics that would otherwise be challenging to study.
    While most open quantum systems studied are modeled as being memory-less (obeying the Markov approximation), real baths generally are influenced by the system-bath interaction, and some systems existing in structured non-Markovian environments can display novel behavior as a result.
    Here we utilize a trapped ion quantum simulator to simulate a single spin-$1/2$ driven-dissipative system with a non-Markovian dissipation channel, and experimentally compare steady-states to those from an analogous Markovian bath.
    We observe that a non-Markovian dissipative channel can shift the steady-state even for a single qubit, to a regime inaccessible for Markovian dissipation.
    % This demonstrates the added richness available to quantum systems in structured environments.
    The techniques used here are compatible with many-body extensions of the model, which can not be simulated efficiently on a classical computer in general. 
    Our work also opens up new possibilities in quantum reservoir engineering beyond the Markovian regime.

\end{abstract}

\maketitle

The study of quantum many-body systems has grown increasingly vital as modern technologies probe regimes where complex quantum effects dominate. 
In realistic physical platforms, dynamics are rarely governed by unitary evolution alone, but also shaped by dissipative interactions mediated by an environmental bath. 
Strikingly, this continuous exchange can drive a system toward novel steady-states that exhibit dissipative phase transitions (DPTs)—phenomena that frequently lack any equilibrium equivalent \cite{DDIM_maghrebi,haack2023probingnonequilibriumdissipativephase}. 
Characterizing these non-trivial dissipative phases is essential both for understanding open quantum many-body systems \cite{Lambert2013} and for engineering fault-tolerant quantum hardware amidst environmental noise \cite{Paladino2014}.

Information entering an ideal bath rapidly spreads across many degrees of freedom, forming a dissipation channel.
However, real baths are finite or structured, such that the rate of information transfer into the bath's degrees of freedom is never truly infinite or constant.
This time-varying, finite-rate coupling is characteristic of non-Markovianity, as the system-bath interaction retains some memory which is not instantaneously erased.
The specific structure of this quantum channel can allow for non-trivial dynamics, potentially triggering DPTs to phases that do not have an analog in Markovian systems \cite{spect_theory_non_markovian}.
Although theoretical approaches using spectral theory can provide some clarity in this regime \cite{spect_theory_non_markovian, spect_theory_liouvillians}, experimental explorations of these systems are lacking.

Trapped ions, with their long coherence times, highly tunable coherent interactions, and accessible dissipative channels; serve as a good platform for probing these driven-dissipative systems.
Specific DPTs have been explored in trapped ions in various experiments \cite{H1_DPT, Cai2022DissipativePhaseTransition}, as have other related studies involving engineered dissipation \cite{Barreiro2011OpenSystemSimulator, Lin2013DissipativeBellState, So2025QuantumSimulation, Kienzler2015ReservoirEngineering}.

Here we focus on engineering a driven-dissipative system with arbitrary bath spectrum for an effective single spin-$1/2$ system formed by a \yb ion. 
This is achieved by simulating many quantum trajectories of the open system, with each trajectory containing multiple quantum jumps interwoven with coherent Hamiltonian evolution. 
%We implement each quantum jump using a deterministic dissipative process, but with the timing of the quantum jump events statistically sampled based on the desired spectral properties of the bath.
%The ensemble average of many quantum trajectories yields identical evolution of the system under an engineered, generally non-Markovian bath.
Our method of engineering dissipation also circumvents technical challenges in realizing controlled dissipation on quantum hardware, most notably preventing leakage to extraneous states. 
%Generally, engineered dissipation is mediated through couplings outside the computational subspace, potentially causing detrimental leakage to extraneous states. 
%In our experiment, dissipation is only implemented when the Hamiltonian evolution is paused, similar to the recently studied Floquet dissipation \cite{Sierant_2022,haack2023probingnonequilibriumdissipativephase}. 
%We can thus use this sequence to simultaneously perform the dissipative interaction and re-pump the qubit back into the computational basis if it escapes.
We explore systems for which the dissipation channel is non-Markovian, and we demonstrate features in the steady-state of a non-Markovian system which are not seen for any Markovian analog.
We use methods which can readily extend to systems with many-body interactions, laying a groundwork for quantum simulations of more complex driven-dissipative systems governed by non-Markovian dissipation.

We start by considering a driven-dissipative single spin-$1/2$ system with amplitude damping and dephasing. 
The Hamiltonian and Lindblad jump operators associated with this model are
\begin{equation}
    H = B \sigma_y,  \ L_1 = \sqrt{\Gamma}\ket{\downarrow}\bra{\uparrow},  \ \mathrm{and} \ L_2 = \sqrt{\Gamma}\ket{\downarrow}\bra{\downarrow},
    \label{eq:Hamiltonian_and_jump_ops}
\end{equation}
where $B$ is the coherent drive strength around the $\mathrm{y}$ axis, and $\Gamma$ is the dissipation rate, and $\ket{\downarrow}$ and $\ket{\uparrow}$ are eigenstates of the $\sigma_z$ operator. 
Dynamics are governed by a Lindblad master equation as 
\begin{align}
    \dot \rho (t)&= \frac{1}{i\hbar}\big[B\sigma_y,\rho(t) \big] + \Gamma \big( \ket{\downarrow}\bra{\downarrow} -\rho(t)  \big),
    \label{eq:MasterEq}
\end{align}
where the collective effect of $L_1$ and $L_2$ results in a simplified dissipator implementing a reset operation to $\ket{\downarrow}$.
We are interested in the features of the steady-state of this system $\rho_{ss} = \lim_{t\rightarrow \infty}\rho(t)$, studying the magnetization orthogonal to the drive $M_x$ and $M_z$, using the notation $\langle M_x\rangle$ or $\langle M_z\rangle$ to specifically denote the steady-state average.
The ratio of dissipation to coherent drive strength $\Gamma/B$ affects the steady-state of this system, as seen in Fig. \ref{fig:unstructured_magnetization_plot_simple}.
% With high $\Gamma/B\gg 1$, frequent resets limit the system's ability to evolve away from its state post-reset (Zeno polarized).
% Alternatively, $\Gamma/B \ll 1$ generates a steady-state with equal scrambling across all states (steady-state depolarized).
%Between these limits at $\Gamma/B=2$, steady-state magnetization associated with the drive is maximized.

Considering only a single quantum trajectory of the ensemble average  $\rho_{ss}$, the moments at which the state resets to $\ket{\downarrow}$ would resemble a Poisson process with rate $\Gamma$.
The likelihood of a quantum jump on a given interval is a function only of that interval's length and the instantaneous state.
We identify this condition as Markovian since the dissipation mechanism has no memory.
% In this particular example, the likelihood of undergoing a state-reset is independent of the momentary quantum state.

Next, we consider a generalization of the above model, where the bath is non-Markovian and the evolution is instead described by a Nakajima-Zwanzig generalized master equation \cite{Breuer2002}, given by 
\begin{equation}
    \dot\rho(t) = -i[H,\rho(t)]+\int_0^t\mathcal{K}(t-t')[\rho(t')]dt'.
    \label{eq:NZ_GME}
\end{equation}
% The memory kernel $\mathcal{K}(t)$ describes how strongly the system's history at time $t'$ influences the state at time $t$. 

Specifically, we define $f(t)$ as the underlying relative dissipative probability distribution in time, i.e. if within a given quantum trajectory a quantum jump (in this case reset to the state $\ket{\downarrow}$) occurs at time $t=t_i$, then $f(t')$ is the probability density of the next quantum jump occurring at $t=t_i+t'$.
Specifying $f(t)$ with the jump operators fully determines the memory kernel $\mathcal{K}(t)$, and encodes the spectral properties of the system-bath interaction (see Appendix A).
We obtain an effective dissipation rate $\Gamma_{\mathrm{eff}}$, defined through
\begin{equation}
    1/\Gamma_{\mathrm{eff}} = \bar{t} = \int_0^\infty tf(t)dt ,
    \label{eq:gamma_effective}
\end{equation}
which is also the inverse mean time $\bar{t}$ between consecutive jump events. 

%In a Markovian system, $f(t) = \Gamma e^{-\Gamma t}$ is an exponential decay with $\Gamma_\text{eff}=\Gamma$, and one can show that the dynamics of Eq.\,\eqref{eq:NZ_GME} reduces to the Markovian dynamics governed by Eq.\,\eqref{eq:MasterEq} (see Appendix A). 
The dynamics for a general $f(t)$ can still reach a steady-state at sufficiently long time, with the time $T$ required to erase the information about the initial state generally satisfying
\begin{equation}
    \bar{N} = \frac{T}{\bar t} \gg \frac{{\bar t}^2}{\sigma^2} ,
    \label{eq:steady_state_condition}
\end{equation}
where $\bar{N}$ is the expected number of dissipation events in time $T$ and $\sigma^2= \bar t^2-\int t^2f(t)dt$ is the variance of $f$.
%We point out that the Zeno-polarized and steady-state depolarized regimes mentioned in the Markovian case still hold for any choice of $f(t)$, although the intermediate crossover between the two regimes can be quite different from the Markovian case.

As an example of a non-Markovian bath, we examine a Pareto-correlated $f(t)$ with a power-law decay in time with exponent $\alpha$, given by
\begin{equation}
f(t) =
\begin{cases}
0 & \text{if } t < \tau \\[6pt]
\dfrac{\alpha \tau^{\alpha}}{t^{\alpha+1}} 
& \text{if } t \ge \tau.
\end{cases}
\end{equation}
This distribution features a delayed recovery time $\tau$ during which further resets are paused. It is also analogous to "colored" noise commonly found in solid-state quantum hardware \cite{Paladino2014}.
Normalization is satisfied for $\alpha>0$, and to satisfy Eq.\,\eqref{eq:gamma_effective}, we set $\tau = \frac{\alpha-1}{\alpha\Gamma_\text{eff}}$.
For the remainder of this work, we will focus on this specific non-Markovian model, though our experiments are adaptable to any arbitrary choice of $f(t)$.
More specifically, we fix $\alpha = 4$ as the shape parameter, chosen as a tradeoff between distinct steady-state features and the experimental runtime required by Eq.\,\eqref{eq:steady_state_condition}. 
%As shown in Fig.\,\ref{fig:shift_Markov_to_nonMarkov}, despite the dissipation resetting the ion to the state in the $-\mathrm{z}$ direction, a positive steady-state magnetization in that direction is obtainable for intermediate values of $\Gamma_\mathrm{eff}/B$. This is in fact impossible to achieve using the Markovian model for any ratio of $\Gamma_\mathrm{eff}/B$ (see Fig.\,\ref{fig:unstructured_magnetization_plot_simple}).
\begin{figure}
    \centering
    \includegraphics[width=1\linewidth]{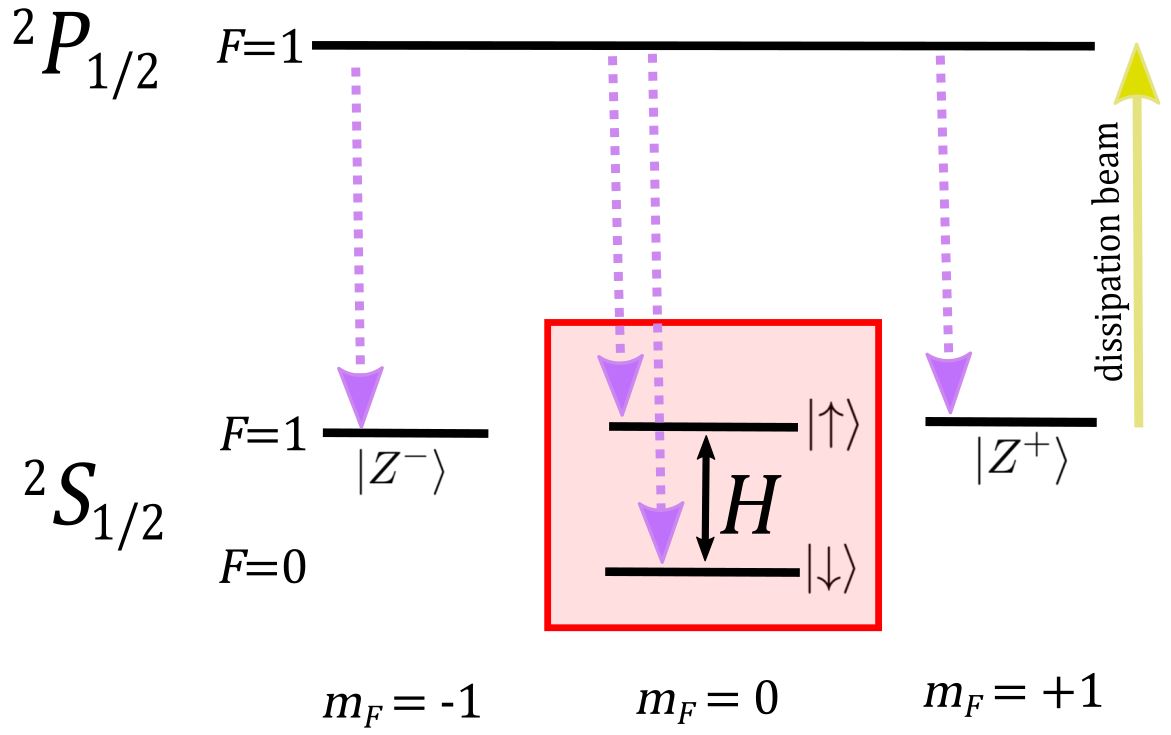}
    \caption{Relevant energy levels in \yb ions.  $\ket{^2S_{1/2},F=0, m_F=0}=\ket{\downarrow}$ and $\ket{^2S_{1/2},F=1, m_F=0}=\ket{\uparrow}$ encode the eigenstates of $\sigma_z$, and span the computational basis (red square).  Optical pumping beam (solid yellow arrow) couples bright states to the excited $^2P_{1/2}$ levels.  Spontaneous decay pathways (purple dotted arrows) enable relaxation to the $\ket{^2S_{1/2},F=1, m_F=\pm1}=\ket{Z^\pm}$ states, which exist outside the computational basis and do not evolve under the coherent drive $H$.}
    \label{fig:decays_Yb_main}
\end{figure}
\begin{figure*}
    \centering
    \includegraphics[width=1\linewidth]{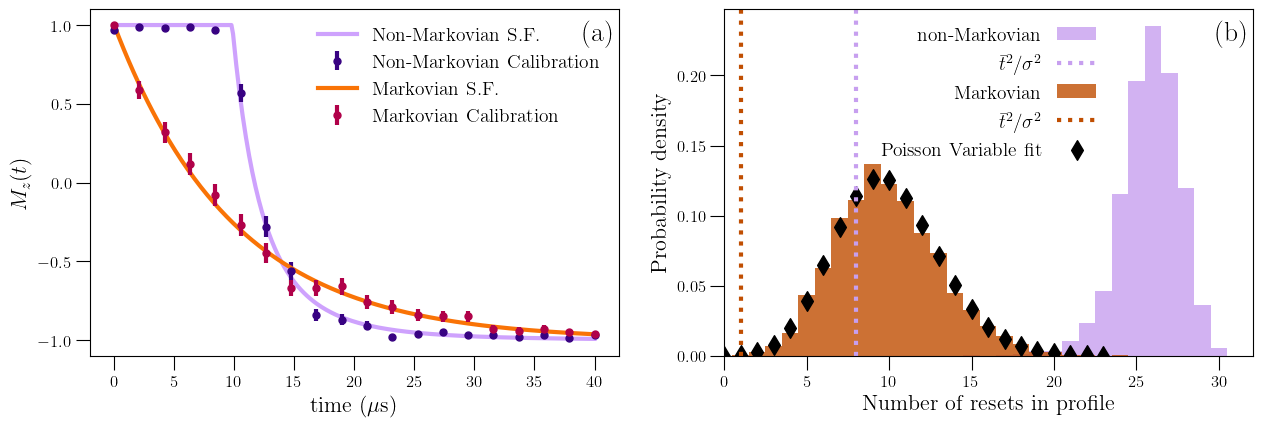}
    \caption{Calibrations of stochastic dissipation procedure.
    (a) The state is initialized to $\ket{\uparrow}$, and then optical pumping is stochastically applied with $\Gamma_\text{eff}\approx100$kHz, reproducing the survival function (SF) of $f$ for both Markovian (exponential) and non-Markovian (Pareto) profiles. 
    (b) Distribution of total resets across profiles.
    Engineered Markovianity agrees with a Poisson process.
    The non-Markovian dissipation used here requires a higher average number of resets due to less variance in $f$.
    The total time $T$ is set long enough to approximate the steady-state (histograms above dotted lines).}
    \label{fig:combined_g_and_number}
\end{figure*}

We trap a \yb ion using a RF-Paul trap, encoding the $\ket{F=0, m_F=0}$ and $\ket{F=1, m_F = 0}$ states of the ground $^2S_{1/2}$ manifold as $\ket{\downarrow}$ and $\ket{\uparrow}$ of an effective spin-$1/2$ system.
$12.642$ GHz radiation from a RF horn antenna drives coherent Rabi oscillations between these states.
The $^2S_{1/2}$ manifold also contains the auxiliary $\ket{F=1, m_F=\pm 1}$ states, which we label $\ket{Z^\pm}$ respectively.
These states, along with $\ket{\uparrow}$, are "bright" as they will fluoresce under a $369.5$nm laser resonant to the roughly cycling $\ket{^2S_{1/2}, F=1}$-$\ket{^2P_{1/2},F=0}$ transition.  
However, $\ket{Z^\pm}$ don't couple to the drive.
Optical pumping beams (state reset) couple all bright states to the $\ket{^2P_{1/2},F=1}$ states, each allowing spontaneous decay to two bright states and the $\ket{\downarrow}$ state.
With enough subsequent scattering events, the state population will eventually reach the "dark" $\ket{\downarrow}$ state.
State discrimination is implemented by imaging state-dependent fluorescence onto a photo-multiplier tube (Hamamatsu: H10682-210) for state readout.

% Spontaneous decay from the $^2P_{1/2}$ states is dipole-allowed to at least one auxiliary level , so for each scattering event, the ion has some likelihood of exiting the qubit manifold.
Although the auxiliary $\ket{Z^\pm}$ levels still couple to the excited state via the pumping beam, the coherent drive only couples the $\ket{\uparrow}$ and $\ket{\downarrow}$ states.
With regions of interest falling near $\Gamma_\mathrm{eff}/B \approx 1$, the scattering timescale from these states is comparable to the characteristic time of the drive, thus skewing the dynamics and breaking $2$-level system equivalence.
Therefore we must limit time spent in auxiliary states through rapid re-pumping into the qubit manifold, requiring additional consideration beyond applying a standard optical pumping beam simultaneously with the drive.
Our solution is to apply strong pumping in a Floquet fashion similar to previous proposals \cite{haack2023probingnonequilibriumdissipativephase}.
By probabilistically applying an operation that deterministically resets the state, across many independent trials, we obtain a collection of "reset profiles" that approximate the ensemble of quantum trajectories available to the model system, without populating the $\ket{Z^\pm}$ states.
% In the model system, quantum trajectories taken over subsequent shots will be independent, so we collect many reset profiles but only perform one experimental shot per profile to better simulate the ensemble.
% This method extends naturally to explorations beyond Markovian dissipation.
By filtering the reset profiles in the ensemble, or by generating reset profiles with correlations between scattering events, non-Markovian dissipation can be simulated.
\begin{figure*}[t]
    \centering
    \includegraphics[width=1\linewidth]{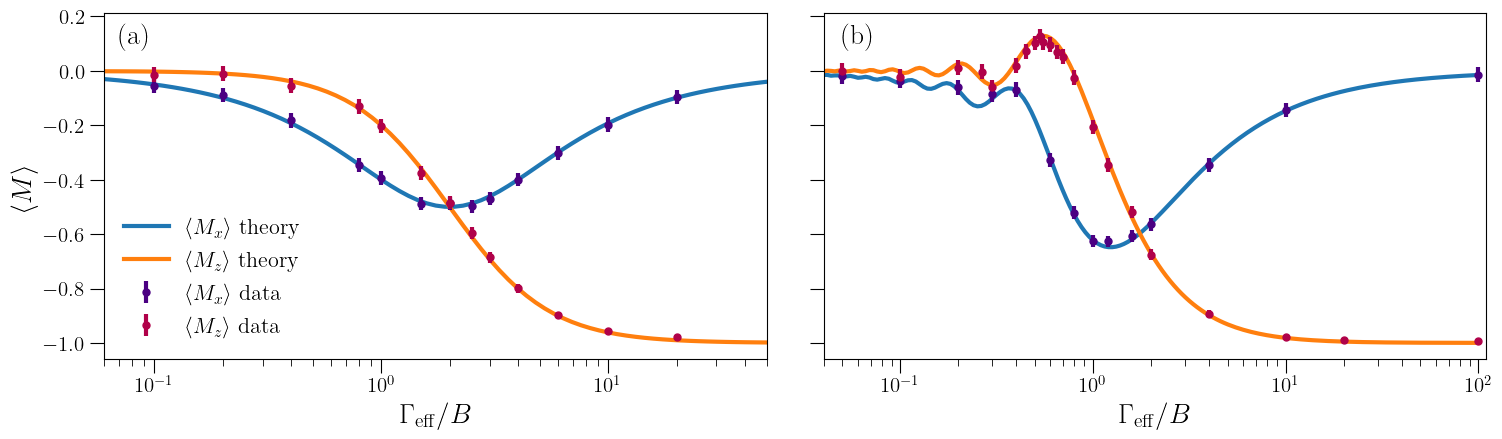}
    \caption{Steady-state magnetization under Markovian (a) and non-Markovian (b) dissipation.  
    Magnetization is measured along $\hat x$ (blue) and $\hat z$ (red) directions, agreeing with theory. 
    Error bars represent shot noise across $10,000$ reset profiles, including SPAM error of $< 1\%$. We find steady-states that are magnetized oppositely to the Zeno polarized state near $\Gamma_\text{eff}/B \approx0.5$, which is unobtainable from systems with Markovian dissipation.
    Only $z$-magnetization sees this effect due to symmetry in Rabi oscillations for $\hat x$.
}
    \label{fig:Markovian_nonMarkovian_bloodstone_apr2026}
\end{figure*}

To that end, we perform the following sequence on our experimental quantum simulator.
We first specify the correlation function $f(t)$, scaled according to $\Gamma_\text{eff}$, and select the total time $T$ according to Eq.\,\eqref{eq:steady_state_condition}.
We obtain random samples $\{ t_i\}$ from $f(t)$, until the total of these samples $\sum\limits_{i=0}^{N} t_i >T$.
At the moments $\{\sum\limits_{i=0}^k t_i \quad \forall k<N\}$, we pause the coherent drive and insert a strong optical pumping sequence which deterministically resets the ion's state.  The strong pumping is applied for $5t_e$, where $t_e$ is the time after which residual population in the bright states falls to $1/e$.
    % The duration of the pumping sequence does not count towards $T$.
We optionally append a qubit rotation to shift the measurement basis, then collect state-dependent fluorescence.
% \begin{enumerate}
%     \item Specify the correlation function $f(t)$, scaled according to $\Gamma_\text{eff}$.
%     Select the total time $T$ according to Eq.\,\eqref{eq:steady_state_condition}.
%     \item Obtain random samples $\{ t_i\}$ from $f(t)$, until the total of these samples $\sum\limits_{i=0}^{N} t_i >T$.
%     \item At the moments $\{\sum\limits_{i=0}^k t_i \quad \forall k<N\}$, pause the coherent drive and insert a strong optical pumping sequence which deterministically resets the ion's state (remaining population in bright states $\approx e^{-5}$).
%     % The duration of the pumping sequence does not count towards $T$.
%     \item Optionally, append a qubit rotation to shift the measurement basis.  Collect state-dependent fluorescence.
% \end{enumerate}
% This sequence faithfully reproduces the ensemble statistics of a system with any arbitrary structure $f$.
We repeat this for approximately $10,000$ reset profiles for each value of $\Gamma_\text{eff}/B$.
Between each point, the instantaneous Rabi frequency $B$ is calibrated so that relative uncertainty in $\Gamma_\text{eff}/B$ remains below $10^{-3}$.
State-preparation and measurement error did not exceed $0.9\%$ throughout data collection.

As a proof of principle, we perform the experiment first by turning off the coherent drive altogether.
By initializing to $\ket{\uparrow}$, we reproduce the survival function of $f$ as shown in Fig. \ref{fig:combined_g_and_number}a.
We also verify the statistics of the scattering events generated after the full experiment length $T$, shown in Fig. \ref{fig:combined_g_and_number}b.

% \begin{figure}
    
%     \includegraphics[width=1\linewidth]{DSSPT_img/Bloodstone_data/Apr2026_double_synthesized_g.png}
%     \caption{Probabilistic operation of a deterministic optical pumping procedure.
%     The state is initialized to $\ket{\uparrow}$, and then optical pumping is applied in the stochastic method using $\Gamma_\text{eff}\approx100$kHz, reproducing the survival function of $f$ for both Markovian (exponential) and non-Markovian (Pareto) profiles.}
%     \label{fig:optical_pumping_g}
% \end{figure}

% \begin{figure}

%     \includegraphics[width=1\linewidth]{DSSPT_img/Bloodstone_data/Apr2026_combined_num_resets_hist.png}
%     \caption{Distribution of total resets across profiles.
%     Engineered Markovianity agrees with a Poisson process.
%     The non-Markovian dissipation in this work requires a higher average number of resets due to less variance in $f$.
%     The total time $T$ is set long enough to approximate the steady-state (histograms above dotted lines).}
%     \label{fig:combined_num_resets_hist}
% \end{figure}

We show experimental data for a system with Markovian dissipation in Fig. \ref{fig:Markovian_nonMarkovian_bloodstone_apr2026}a.
The coherent drive strength is not modified between points, but rather $\Gamma_\text{eff}$ (through $f$) scales to probe different relative dissipation strengths.
To maintain the steady-state approximation equally across all data points, the final $T$ is adaptively changed so the mean number of dissipation events per profile is fixed as $\Gamma_\text{eff}$ varies, and falls well beyond the steady-state condition (Fig. \ref{fig:combined_g_and_number}b) $\bar N=T/\bar t \approx 10 \gg \bar t^2/\sigma^2 = 1$.
% The probability of no dissipation events occurring by the final time $T$ is below $5\times10^{-5}$, well below other errors.
% SPAM error was calibrated once every two hours, and did not exceed $0.9\%$ throughout data collection.

We clearly identify the Zeno polarized and steady-state depolarized regimes, and match with predicted values (see Appendix A).
This agreement also demonstrates that the $\ket{Z^\pm}$ populations were properly extinguished, making this a faithful simulation of the $2$-level model.

For non-Markovian dissipation (Fig. \ref{fig:Markovian_nonMarkovian_bloodstone_apr2026}b)
the experiment time $T$ was longer to satisfy the steady-state condition $\bar N=T/\bar t \approx 26 \gg \bar t^2/\sigma^2 = 8$, owing to the decreased relative variance.
At intermediate dissipation strength, we observe steady-states oppositely magnetized to the Zeno state, which cannot be seen at any relative dissipation strength for a Markovian dissipation channel (see Appendix A).
We measure $\langle M_z\rangle(\Gamma_\text{eff}/B=0.53) = 0.13 \pm0.03$ as the maximal point.
This only occurs in the direction of the resets due to symmetry of the Rabi oscillations, which we confirm with data collected for $\langle M_x\rangle$.

%\iffalse
%with 
%$\langle M_z\rangle(\Gamma_\text{eff}/B=0.55) = 0.10 \pm0.03$ and $\langle M_z\rangle(\Gamma_\text{eff}/B=0.6) = 0.10 \pm0.03$.
%\hl{uncertainty in $\Gamma_\text{eff}/B$ is approximately $3E-4$ relative error based on pi time calibrations (fit error and drift).}
%\fi

% \begin{figure}
%     \centering
%     \includegraphics[width=1\linewidth]{DSSPT_img/Bloodstone_data/Apr2026_nonMarkov_figure.png}
%     \caption{Steady-state magnetization expected vs experimental for a non-Markovian dissipation channel defined by a Pareto ($\alpha=4$) distribution.  }
%     \label{fig:nonMarkovian_bloodstone_apr2026}
% \end{figure}

In summary, we demonstrated quantum simulation of steady-states in a single spin-$1/2$ driven-dissipative system under both Markovian and non-Markovian dissipation channels. 
%We performed an ensemble average to recover the effective $2$-level dynamics from a $4$-level physical system represented by the internal energy levels of a trapped ion, avoiding leakage to auxiliary bright states.
We observed new features in the steady-states of the qubit which arise from non-Markovianity, including steady-state observables unobtainable for any dissipation strength in a Markovian system.

While applying multiple resets in a single-body system is technically redundant, the experimental techniques used here are readily adaptable to many-body systems, in which the entire history of quantum jumps is important since not all spins/qubits jump simultaneously in general. 
Stimulated Raman transitions may be used in trapped ions to generate the many-body coherent couplings of such models \cite{Molmer1999Multiparticle}.
Independent probabilistic resets on each ion would require efficient mid-circuit individual-qubit operations.
Recent experiments have demonstrated such capabilities in trapped ions through holographic beam shaping \cite{Motlakunta2024Preserving} and other techniques \cite{Yu2025MidCircuitReset}.

In these many-body cases, non-Markovianity in the dissipation channel may give rise to distinct behavior unobtainable from Markovian dissipation. Studying such systems using a quantum simulator is highly advantageous, as the exponentially growing Hilbert space combined with the need to track the entire history of quantum trajectories makes classical simulation difficult.
In these more complex systems, the applicable `structure' for non-Markovian dissipation can correspond not only to temporal but also spatial correlations in the dissipation channel. In fact, spatial correlations in the dissipation naturally exist in trapped ion platforms due to the crosstalk of lasers addressing nearby ions. Beyond measuring features of the steady-states, the experimental methods developed here can also be adapted to investigate transient dynamics following a quench of either the Hamiltonian or dissipation. Such transient dynamics are even harder to compute classically with non-Markovian baths.

Our work lays an experimental framework for quantum simulations of driven-dissipative quantum many-body systems with non-Markovian baths, or dissipative quantum state preparation that may benefit from non-Markovian dissipation \cite{Seki2026Dissipative}.
In the limit of nearly perfect periodicity in the dissipation channel, one could also explore dissipative discrete time crystals with a non-Markovian bath, which may lead to new physics beyond those already being studied \cite{Camacho2024QuantumClassicalFeedback,Camacho2026TimeCrystals, Zhang2017TimeCrystal}.
Finally, the Pareto distribution studied in this work may be useful in studying the $1/f$ noise common in solid-state quantum hardware, and could possibly be further explored to improve the performance of solid-state qubits with a better understanding of their environment \cite{Paladino2014}.

\section{Acknowledgments}

We acknowledge financial support from the Canada First Research Excellence Fund (CFREF), the Natural Sciences and Engineering Research Council of Canada (NSERC) Discovery program (RGPIN-2018-05250 and RGPIN-2025-06496) and Alliance grants (ALLRP 588338 - 23 and ALLRP 592646 - 24), University of Waterloo, and The Strategic Science Fund of the Government of Canada.
The authors thank Siddharth Chawla for contributions to preliminary experiments.  The authors also thank Mahmoud Badawy, Chris Graham, and Faizan Samad for help with the experimental control system.

\bibliographystyle{apsrev4-2}
\bibliography{references.bib}

\appendix
\clearpage
\section{Appendix}
\subsection{A: Generalized Master Equation}

The memory kernel $\mathcal{K}(t)$ used in Eq.\,\eqref{eq:NZ_GME} can be decomposed as 
\begin{equation}
    \mathcal{K}(t-t')[\rho(t')] = k(t-t')\big(\ket{\downarrow}\bra{\downarrow}-\rho(t')\big)
    \label{eq:memory_kernel}
\end{equation}
where $k$ is a scalar memory function which determines the variation in dissipation strength \cite{Breuer2002}.
The scalar memory function $k(t)$ is defined using the Laplace transform of the relative temporal dissipative probability $f(t)$ according to
\begin{equation}
    \widetilde{k}(s) = \frac{s\widetilde{f}(s)}{1-\widetilde{f}(s)} \ \ .
    \label{eq:Laplace_transform}
\end{equation}

In the Markovian limit, we have $f(t) = \Gamma e^{-\Gamma t}$, implying $\widetilde{f}(s) = \frac{\Gamma}{s+\Gamma}$.
The scalar kernel in the Laplace domain then becomes $\widetilde{k}(s) = \Gamma$, such that the inverse Laplace transform is $k(t) = \Gamma \delta(t)$.
The Dirac delta function collapses the integration in Eq.\,\eqref{eq:NZ_GME}, resulting in the master equation in Eq.\,\eqref{eq:MasterEq}.

A similar analysis may be performed in order to determine the $f(t)$ required to simulate a bath with specific spectral properties.

The Generalized Master equation is not easy to solve for arbitrary times, however it is easier to solve for the steady-state limit for a single spin.
We label the survival function $g(t) = 1-\int^{t}_0f(t')dt'$ as the probability that no quantum jump occurs by time $t$.
Let $h(t)$ be the renewal density function \cite{Cox1962,Piilo2008NonMarkovian}, which represents the absolute probability density that a jump occurs exactly at time t, regardless of the prior jump history.  
At an arbitrary moment $T$, the state density is easily identified as 
\begin{align*}
\rho(T) = &g(T)U(T)\rho(0)U^\dagger(T) \\&+ \int_0^Th(T-t)g(t)U(t)\ket{\downarrow}\bra{\downarrow}U^\dagger(t)dt
    \label{eq:renewal_density}
\end{align*}
where $U(t) = \mathrm{exp}(-iHt)$ is the operator for pure coherent evolution.
The first term encodes the trajectories which have no quantum jumps, and the second term encodes those which have at least one quantum jump.  
In many-body cases, one would need to track not just the second term but the whole set of all combinations of jumps owing to the residual information which remains after a partial state-reset.
For the steady-state, we are able to ignore the first term because of the normalization condition, and the second term simplifies 
\begin{equation}
\lim_{T\to\infty} g(T) =0 \quad,\quad \lim_{T\to\infty} h(T) = \Gamma_\mathrm{eff} = 1/\bar t
\end{equation}
owing to the constancy of the renewal function in the long time limit.  
From here, the above equation simplifies greatly, to
\begin{equation}
    \lim_{T\to\infty}\rho(T) = \rho_{ss}=\int_0^\infty \Gamma_\mathrm{eff} g(t)\big( U(t)\ket{\downarrow}\bra{\downarrow}U^\dagger(t)\big)dt
\end{equation}
and therefore the expectation of any steady-state observable $\hat O$ is given by 
\begin{equation}
    \langle \hat O\rangle = \int_0^\infty \Gamma_\mathrm{eff} g(t)\mathrm{Tr}\bigl[\hat O U(t)\ket{\downarrow}\bra{\downarrow}U^\dagger(t)\bigr]dt
\end{equation}

With the Hamiltonian in Eq.\,\eqref{eq:Hamiltonian_and_jump_ops}, and looking at the magnetization in the other Bloch axes, this produces
\begin{align}
    \mathrm{Tr}\bigl[\hat M_x U(t)\ket{\downarrow}\bra{\downarrow}U^\dagger(t)\bigr] &= -\mathrm{sin}(2Bt) \\
    \mathrm{Tr}\bigl[\hat M_z U(t)\ket{\downarrow}\bra{\downarrow}U^\dagger(t)\bigr] &= -\mathrm{cos}(2Bt)
\end{align}
and the inner product of these terms with the survival function $g(t)$ produces the expectation curves used throughout this work.

For the special case of Markovian dissipation with $\mathrm{z}$-magnetization, we have
\begin{equation}
    \langle M_z\rangle = -\int^\infty_0 \Gamma e^{-\Gamma t}\cos (2Bt)dt = -\frac{\Gamma^2}{\Gamma^2+4B^2}
\end{equation}
which is strictly negative for real $\Gamma$.
Therefore to equate the positive $\mathrm{z}$-magnetization found in the non-Markovian experiments to an "effective" equivalent Markovian experiment would require an imaginary dissipation rate. For eample, to reach a steady-state average magnetization of $\langle M_z\rangle \approx 0.13$ would require an equivalent dissipation strength of approximately $\Gamma/B \approx 2i/3$.

\begin{figure*}[t]
    \centering
    \includegraphics[width=1\linewidth]{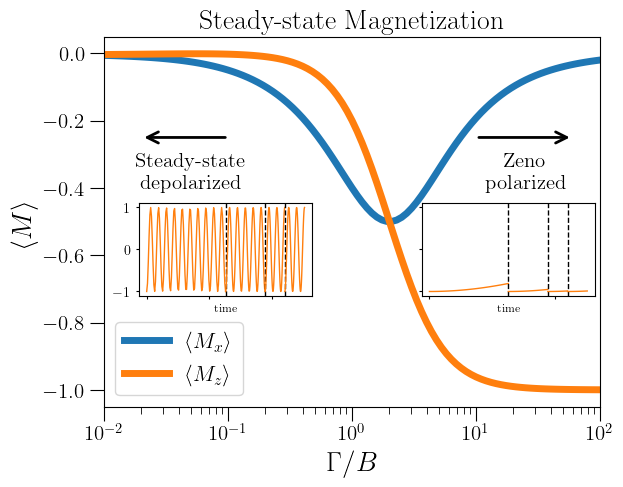}
    \caption{Expected steady-state magnetization as reset strength $\Gamma$ changes relative to drive strength $B$ for model in Eq.\,\eqref{eq:MasterEq}.
    % Maximum coherent buildup is at $\Gamma/B = 2$.  
    Resetting too infrequently leads to scrambling of final evolution length (steady-state depolarized).  Resetting too frequently limits coherent evolution, (Zeno polarized).  
    Insets show example quantum trajectories for $\mathrm{z}$-magnetization as these limits are approached.}
    \label{fig:unstructured_magnetization_plot_simple}
\end{figure*}

\begin{figure*}[t]
    \centering
    \includegraphics[width=0.7\linewidth]{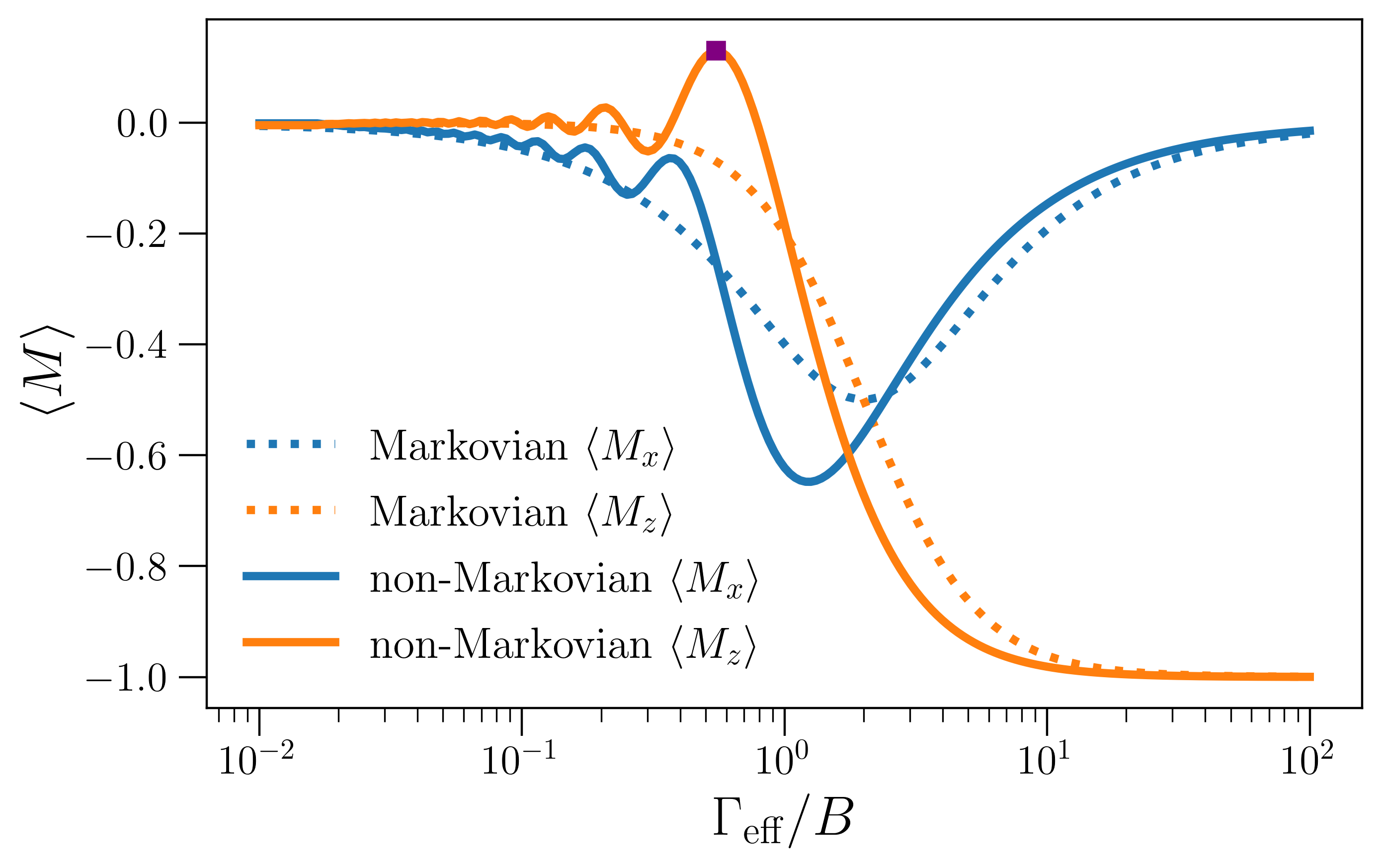}
    \caption{Shift in steady-state magnetization induced by non-Markovianity in the dissipation mechanism, specifically for a Pareto ($\alpha=4$)-correlated dissipation mechanism.  For non-Markovian dissipation, steady-state magnetization can be opposite the reset direction (purple square) for certain intermediate values of $\Gamma_\text{eff}/B$.}
    \label{fig:shift_Markov_to_nonMarkov}
\end{figure*}

\begin{figure*}[t]
    \centering
    \includegraphics[width=1\linewidth]{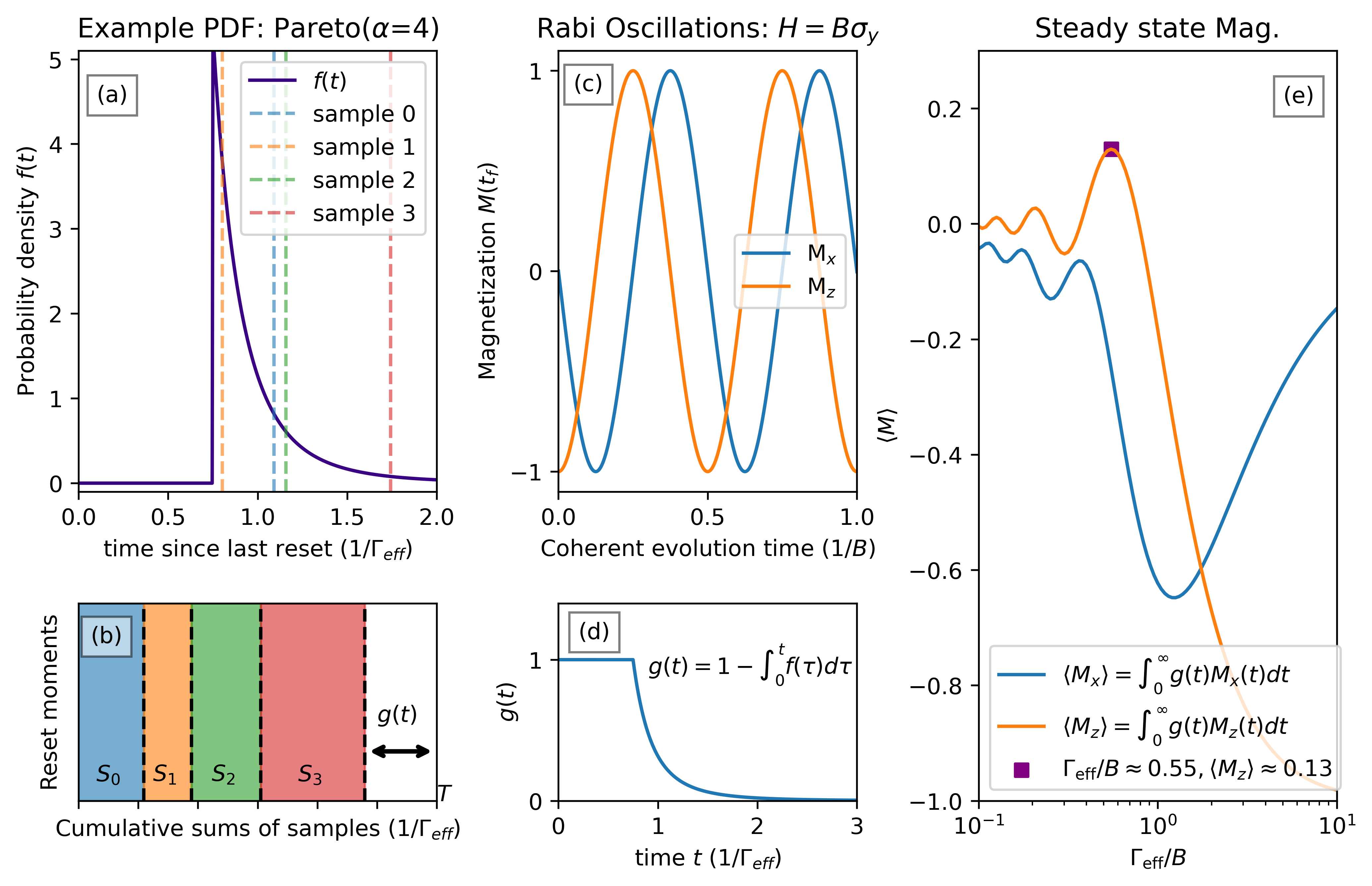}
    \caption{Procedure to implement steady-state experiment for arbitrarily structured dissipation invoking the framework of quantum trajectories. \textbf{(a)} The probability density $f$ determines the time correlations between discrete dissipation events (resets) in the quantum trajectory picture.  Samples can be taken from $f$ (colored dashed lines) to construct an experimental sequence according to these correlations.  \textbf{(b)} For each sample from $f$, apply coherent drive (color) then state-reset (black dashed lines). Samples are iteratively concatenated until the total time goes beyond $T$ chosen such that the state-density closely approximates the steady-state (Eq.\,\eqref{eq:steady_state_condition}).  In the single ion case, this converts $f$ into a distribution $g$ of the final coherent drive period, since any state information is erased. \textbf{(c)} After a reset, the spin undergoes Rabi oscillation with period $(1/B)$.  \textbf{(d)} The distribution $g$ is the survival function of $f$.  Repeating for many independent samples as in (b) should recreate the ensemble average over quantum trajectories $g$, as shown in Fig. \ref{fig:combined_g_and_number}b.
    \textbf{(e)} The steady-state magnetization will be an inner product of the Rabi oscillations with weights according to $g$.  Sweeping $\Gamma_\mathrm{eff}$ or $B$ will result in different relative dissipation strength.  For non-Markovian dissipation, steady-state magnetization can be tuned to be opposite the reset direction (purple square).
    }
    \label{fig:structured_uniform_dissipation_pareto4}
\end{figure*}

\begin{figure*}[t]
    \centering
    \includegraphics[width=0.7\linewidth]{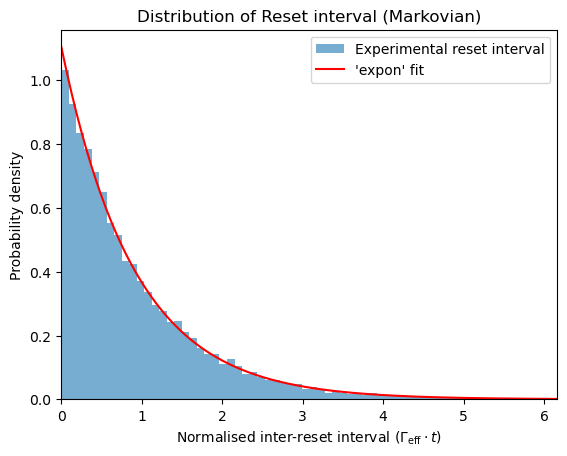}
    \caption{Histogram of the coherent evolution time between resets $\{ t_i\}$ used during Markovian experiments, normalized to the effective dissipation rate $\Gamma_\text{eff}$.  Samples approximate $f$ (an exponential distribution), confirmed with fit to Scipy function "expon". }
    \label{fig:Markov_interreset_interval_hist}
\end{figure*}

\begin{figure*}[t]
    \centering
    \includegraphics[width=0.7\linewidth]{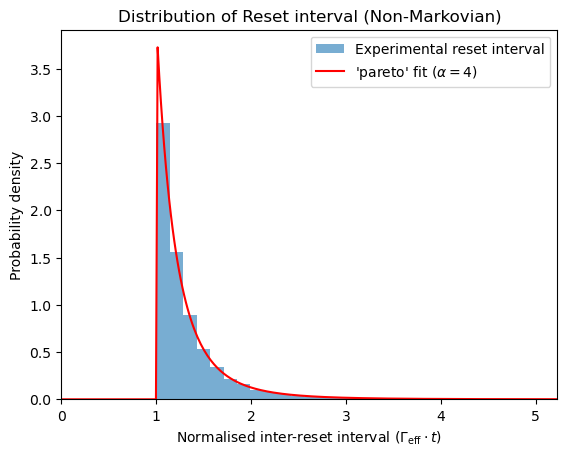}
    \caption{Histogram of the coherent evolution time between resets $\{ t_i\}$ used during non-Markovian experiments, normalized to the effective dissipation rate $\Gamma_\text{eff}$. Samples approximate $f$ (a Pareto distribution with shape $\alpha=4$), confirmed with fit to Scipy function "pareto".}
    \label{fig:nonMarkov_interreset_interval_hist}
\end{figure*}

\end{document}